\documentclass[epj,twocolumn]{webofc}
\usepackage[varg]{txfonts}   
\usepackage{graphicx}

\wocname{epj}
\woctitle{Seismology of the Sun and the Distant Stars 2016}
\begin{document}
\title{The BRITE-Constellation Nanosatellite Space Mission And Its First Scientific Results\thanks{Based on data collected by the BRITE
Constellation satellite mission, designed, built, launched, operated and 
supported by the Austrian Research Promotion Agency (FFG), the University
of Vienna, the Technical University of Graz, the Canadian Space Agency 
(CSA), the University of Toronto Institute for Aerospace Studies (UTIAS),
the Foundation for Polish Science \& Technology (FNiTP MNiSW), and
National Science Centre (NCN).}}
%

\author{\firstname{G.} \lastname{Handler}\inst{1}\fnsep\thanks{\email{gerald@camk.edu.pl}} \and
        \firstname{A.} \lastname{Pigulski}\inst{2} \and
        \firstname{W. W.} \lastname{Weiss}\inst{3} \and 
        \firstname{A. F. J.} \lastname{Moffat}\inst{4} \and
        \firstname{R.} \lastname{Kuschnig}\inst{3,5} \and
        \firstname{G. A.} \lastname{Wade}\inst{6} \and
        \firstname{P.} \lastname{Orlea{\'n}ski}\inst{7} \and
        \firstname{S. M.} \lastname{Ruci\'nski}\inst{8} \and
        \firstname{O.} \lastname{Koudelka}\inst{5} \and
        \firstname{R.} \lastname{Smolec}\inst{1} \and
        \firstname{K.} \lastname{Zwintz}\inst{9} \and
        \firstname{J. M.} \lastname{Matthews}\inst{10} \and
        \firstname{A.} \lastname{Popowicz}\inst{11} \and
        \firstname{D.} \lastname{Baade}\inst{12} \and
        \firstname{C.} \lastname{Neiner}\inst{13} \and
        \firstname{A. A.} \lastname{Pamyatnykh}\inst{1} \and
        \firstname{J.} \lastname{Rowe}\inst{14} \and
        \firstname{A.} \lastname{Schwarzenberg-Czerny}\inst{1}
}

\institute{Nicolaus Copernicus Astronomical Center, Bartycka 18, 00-716 Warsaw,
Poland
\and
Instytut Astronomiczny, Uniwersytet Wrocławski, ul. Kopernika 11, 51-622 Wroc\l{}aw, Poland
\and
Institute for Astrophysics, Universit\"at Wien, T\"urkenschanzstrasse
17, A-1180 Wien, Austria
\and 
D\'epartment de physique, Universit\'e de Montr\'eal, C. P. 6128, Succ. Centre-Ville, Montr\'eal, QC H3C 3J7, Canada
\and
Institut f\"ur Kommunikationsnetze und Satellitenkommunikation, Inffeldgasse 12/I, 8010 Graz, Austria
\and
Department of Physics, Royal Military College of Canada, PO Box 17000, Stn Forces, Kingston, ON K7K 7B4, Canada
\and
Centrum Bada{\'n} Kosmicznych, Polska Akademia Nauk, Bartycka 18A, 00-716 Warszawa, Poland
\and
Department of Astronomy and Astrophysics, University of Toronto, 50 St. George Street, Toronto, ON M5S 3H4, Canada
\and
Institut f\"ur Astro- und Teilchenphysik, Universit\"at Innsbruck, Technikerstrasse 25/8, 6020, Innsbruck, Austria
\and
Department of Physics and Astronomy, University of British Columbia, Vancouver, BC V6T1Z1, Canada
\and
Silesian University of Technology, Institute of Automatic Control, Gliwice, Akademicka 16, Poland
\and
European Organisation for Astronomical Research in the Southern Hemisphere, Karl-Schwarzschild-Str. 2, 85748 Garching b. M\"unchen, Germany
\and
LESIA, Observatoire de Paris, PSL Research University, CNRS, Sorbonne Universit\'es, UPMC Univ. Paris 06, Univ. Paris Diderot, Sorbonne Paris Cit\'e, 5 place Jules Janssen, F-92195 Meudon, France
\and
Institut de recherche sur les exoplan\'etes, iREx, D\'epartement de physique, Universit\'e de Montr\'eal, Montr\'eal, QC, H3C 3J7, Canada
          }

\abstract{%
The BRIght Target Explorer (BRITE) Constellation is the first nanosatellite 
mission applied to astrophysical research. Five satellites in low-Earth 
orbits perform precise optical two-colour photometry of the brightest 
stars in the night sky. BRITE is naturally well 
suited for variability studies of hot stars. This contribution describes 
the basic outline of the mission and some initial problems that needed to 
be overcome. Some information on BRITE data products, how to access 
them, and how to join their scientific exploration is provided. Finally, 
a brief summary of the first scientific results obtained by BRITE is given.

}
\maketitle
%
\section{Introduction}
\label{intro}

A BRIght Target Explorer (BRITE) is a nanosatellite designed as a cube 
of $20 \times 20 \times 20$ cm edge length that weighs approximately 7 
kg. It carries a 3-cm telescope with a wide field of view (about 24{\tt 
$^{\rm o}$} on the sky, nearly unvignetted) that feeds an uncooled $4008 
\times 2672$-pixel KAI-11002M CCD. As such, it is predestined to 
obtain high-precision space photometry of apparently bright stars.

There is more than one BRITE. Each of the partner countries 
participating in the mission, Austria, Canada, and Poland, funded two 
satellites. One of each pair is equipped with a blue-sensitive, the 
other with a red-sensitive filter. The passbands are close to 
Johnson-Cousins $B$ and $R_c$. The whole ensemble of satellites is 
called BRITE-Constellation and is therefore capable of multicolour 
time-resolved photometry.

The six satellites were launched into orbit between February 2013 and 
August 2014. One of the Canadian satellites, BRITE-Montreal, went astray 
during launch as it apparently did not separate from the last stage of 
the rocket. The remaining five satellites are called BRITE-Austria (BAb) 
and Uni-BRITE (UBr, both Austrian), BRITE-Lem (BLb) and BRITE-Heweliusz 
(BHr, both Polish), and BRITE-Toronto (BTr, Canadian), and are 
operating. The abbreviations of the satellites' names, that will be used 
in what follows, originate from their designations followed by the 
filter they carry, red (r) or blue (b).

More detailed accounts about the basic goals and outline of the 
BRITE-Constellation mission have been given by \cite{Weiss14} and 
\cite{Pablo16}. In this contribution, we briefly summarize these 
articles and the first science results, but also provide a glimpse 
into BRITE operations for prospective users.

\section{Scientific goals and observations}
\label{sgoal}

As mentioned in Sect.~\ref{intro}, BRITE-Constellation obtains 
high-precision time-resolved two-colour photometry of bright stars. 
Apparently bright stars are often intrinsically luminous, which means 
that the stellar sample accessible to BRITE contains an 
overrepresentation of hot and evolved stars. As hot stars are 
short-lived, they are concentrated near the Galactic plane. This area is 
of prime interest for BRITE observations, because many target stars and 
bright guide stars used for tracking can be acquired. The primary 
science goal of the mission is thus easy to define: to study the variability 
of hot stars, be it caused by pulsation, rotation, wind variations, 
star-disk interactions, binarity etc. An HR Diagram of the stars 
observable by BRITE is shown in \cite{Weiss14}.

The observing strategy itself bears in mind that the longer the time 
base of observations, the more information on stellar variability can be 
obtained. Consequently, each BRITE observing run attempts to observe a 
given field as long as possible, for up to half a year. This goal is met 
in most cases. Given the low-Earth orbit of the satellites, occultations 
of the target field by the Earth or stray light imply that only a 
fraction of the $\sim 100$-min orbital period (up to 40\%, depending on 
the satellite) can be used for observations.

The ensemble of satellites can be used to observe one field with all 
five of them (diminishing aliases at the satellites' orbital periods), 
or their number is split between two fields. A given satellite may also 
observe more than one field during its orbit. In the first few years of 
operation, it has been found most useful that two pairs of satellites 
(one with a blue filter, the other with a red one) each observe two 
different fields separated by about 6 h in right ascension, with the 
remaining fifth satellite used as a supplement.

\begin{figure*}[htb!]
\centering
\includegraphics[width=\hsize,clip]{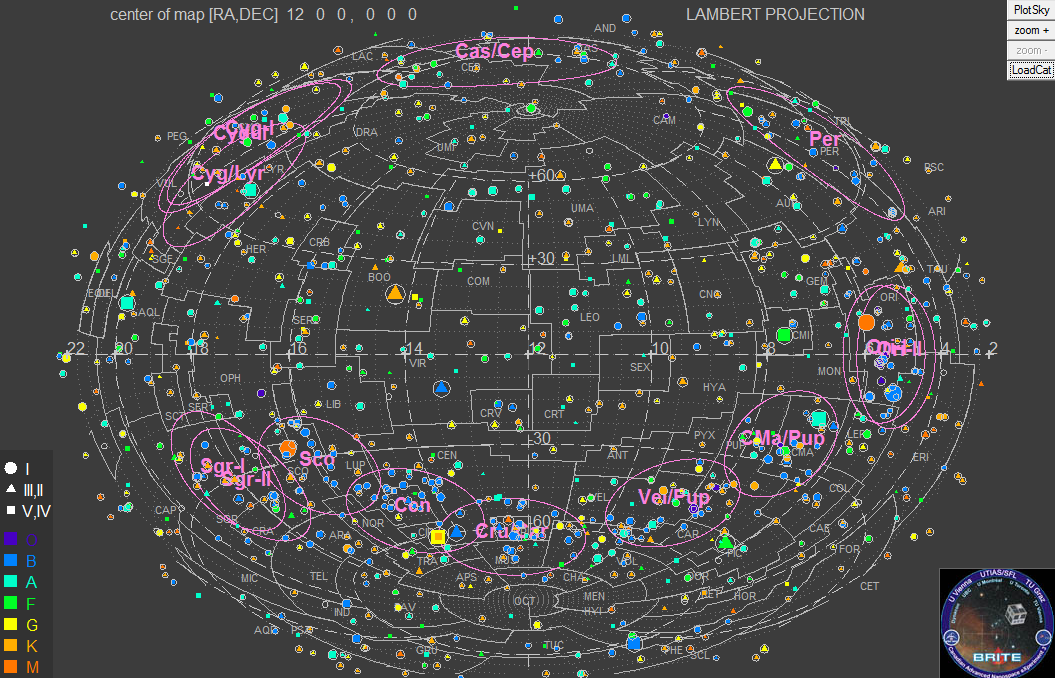}
\caption{A sky map containing all stars brighter than $V=4.5$. The 
colour code refers to spectral type, the symbols to luminosity class. 
Encircled stars have been proposed for BRITE observations. The sky areas 
labeled and encircled in pink have been observed by BRITE to date (cf. 
Table~\ref{tab-obs}). Almost the whole galactic plane has been sampled.}
\label{fig-obs}       
\end{figure*}

The measurements of BRITE-Constellation are being retrieved by three 
ground stations, one in Graz (Austria), one in Warsaw (Poland), and the 
third in Toronto (Canada). Each country operates their satellites 
largely independently, with occasional help by the partners if need be. 
The downlink capacity is limited (also because of some interference 
issues at the European ground stations, see \cite{Pablo16}), and 
therefore full-frame images are generally not downloaded. Instead, 
windows that contain target stars and a sufficiently large area of 
blank sky around them are downloaded for further processing. An overview 
of the BRITE observations so far is presented in Table~\ref{tab-obs}, 
and the observed fields are depicted on a sky map in 
Figure~\ref{fig-obs}.

\begin{table}
\vspace{30pt} 
\caption{Target fields observed by BRITE-Constellation so far. Data 
taken from {\tt http://brite.craq-astro.ca/doku.php}. The total number 
of stars observed is not the sum of the individual ones as some 
stars/fields have been re-observed.}
\label{tab-obs}       
\begin{tabular}{lccrr}
\hline
Field & from & to & $\Delta$T & \#stars\\
 & civil date & civil date & (d)\\\hline
Ori I & 01.12.2013 & 18.03.2014 & 108 & 15\\
Cen & 25.03.2014 & 18.08.2014 & 147 & 32\\
Sgr I & 29.04.2014 & 06.09.2014 & 42 & 19\\
Cyg I & 12.06.2014 & 25.11.2014 & 167 & 36\\
Per & 02.09.2014 & 18.02.2015 & 170 & 37\\
Orion II & 24.09.2014 & 17.03.2015 & 175 & 35\\
Vel/Pup & 11.12.2014 & 28.05.2015 & 169 & 52\\
Sco & 18.03.2015 & 31.08.2015 & 167 & 26\\
Cyg II & 01.06.2015 & 25.11.2015 & 178 & 34\\
Cas/Cep & 23.08.2015 & 17.10.2015 & 55 & 25\\
CMa/Pup & 18.10.2015 & 14.04.2016 & ~180 & 32\\
Cru/Car & 22.01.2016 & 22.07.2016 & 183 & 40\\
Sgr II & 15.04.2016 & 23.09.2016 & 162 & 17\\
Cyg/Lyr & 28.04.2016 & 21.10.2016 & 177 & 15\\\hline
Total & & & & 350\\\hline
\end{tabular}
\end{table}

As soon as the first science images had been obtained by the satellites 
launched earliest, it became clear that the CCDs suffered from radiation 
damages, causing hot pixels and columns, as well as Charge Transfer 
Inefficiency (CTI). Obviously, these defects would have a negative 
influence on the quality of the photometry to be obtained and needed to 
be mitigated. This was accomplished in three ways. First, some of the 
satellites were still on Earth when the problem was noticed. Therefore, 
BHr and BTr were equipped with better shielding, strongly decreasing the 
radiation damage. Second, a sophisticated data reduction pipeline has 
been developed that eliminates these adverse effects to a large extent 
\cite{popo16}.

Third, and most importantly, a ``nodding'' technique has been 
implemented in the observations, starting from early 2015. The BRITE 
satellites no longer stare at the same position. Instead, they are moved 
back and forth by about 0.2$^{\rm o}$ every $\approx 20$~s. Thus, the 
stellar images fall onto alternatingly different parts of their CCD 
windows, whereas the compromised pixels remain the same over such short 
time intervals. Consequently, differential images are computed, which 
are almost free of artifacts, and precise photometry can be obtained 
from these images. We refer to \cite{Pablo16} for more information and 
illustrations of the process. Nodding mode is now the standard observing 
mode for BRITE. It also guarantees that the Constellation can be 
operated for much longer than the projected lifetime of two years 
because radiation damage of the CCDs is now a minor issue in this 
regard.

\section{BRITE data products and policies}
\label{data}

After the CCD data have been processed through the BRITE photometry 
pipeline \cite{popo16}, \cite{popop16}, the data for individual stars 
are assigned to a so-called ``contact PI" who is responsible for their 
final reduction, analysis and publication. This assignment is based on 
calls for proposals in the years 2008, 2011 and 2016 (still open at the 
time of writing, see {\tt http://www.brite-constellation.at/}).

If a given star has been proposed for observation by several 
researchers, these decide between themselves who will become contact PI. 
A contact PI is expected to provide a first report on data analysis 
three months after receiving the data (including a judgement whether or 
not the respective data set is suitable for stand-alone publication), 
and a final report after one year. After this one-year proprietary 
period has passed, the data will become public, unless an extension of 
the proprietary period of up to half a year has been approved in 
exceptional cases. A list of fields observed, stars measured therein, 
the respective contact PIs, and links to public data can be found at 
{\tt http://brite.craq-astro.ca/doku.php}. Even if a data set is still 
in the proprietary phase, the contact PI can be approached for possible 
collaborations, as each contact PI is free to set up their own network 
of co-workers.

Light curves are supplied to scientists in the form of ASCII files. In 
the current format, these contain the timing in HJD, the measured target 
flux, and several additional parameters needed to decorrelate the light 
curves from remaining instrumental effects. These comprise the x/y 
position of the barycentre of the stellar PSF, the CCD temperature, and 
three further parameters describing the shape of the PSF (explained in 
detail by \cite{popop16}).

Decorrelation of the BRITE photometry is required because the data 
orginate from an uncooled instrument (CCD and optics) and there is no 
possibility to acquire calibration images, such as bias, dark and flat 
field frames. Examples of the temperature changes occuring during a run 
on a given field, and within the orbit, are again given by 
\cite{Pablo16}. Decorrelation with temperature is usually most 
important, as the individual CCD pixels' sensitivities depend on it, 
followed by decorrelation with x/y position to compensate for pixel-by-pixel 
sensitivity variations. Since temperature changes also change the PSF, 
decorrelation with the three PSF-related parameters may also be necessary, 
and sometimes the decorrelations must be performed in data subsets 
depending on average CCD temperature. In rare cases, stray light 
corrections need to be applied as well.

Several different approaches have been developed to carry out this 
decorrelation (e.g., \cite{pigu16}, \cite{bram16}). Sometimes it is 
necessary to first remove a target's variability from the observations 
to decorrelate optimally \cite{hand16}). Even though these procedures 
may appear difficult, and several iterations may be required, the 
problems are understood and improved algorithms are permanently being 
developed.

Regarding the organization of the project, the scientific decisions such 
as target field selection, proposal evaluation, proprietary period 
extensions etc., are made by the BRITE Executive Science Team (BEST) 
whose current representatives are also the authors of the present 
manuscript. The scientific user community is organized as the 
BRITE-Constellation International Advisory Team (BIAST). Membership is 
granted after an informal request to one of us (GAW, {\tt 
Gregg.Wade@rmc.ca}) and approval by BEST. BIAST members are informed 
about the actual status of the mission via a monthly email newsletter. 
Last but not least, the BRITE-Constellation Ground-Based Observing Team 
(GBOT) provides a platform for BRITE scientists and observers worldwide 
to support collaborations and maximize the scientific output of 
BRITE-Constellation. The brightness of the targets facilitates 
ground-based support observations: even amateurs with modest-sized 
telescopes have obtained science-quality high-resolution spectroscopy.

\section{First scientific results}

The first three scientific papers have been published in a short series 
\cite{pigu16}, \cite{weiss16}, \cite{baade16}, and have examined three 
different types of object in the Centaurus field. At this early stage in 
the mission, only the Austrian BAb and UBr were available in full 
science mode, and were later joined by BLb and BTr after their 
commissioning had been completed.

First, the brightest rapidly oscillating Ap (roAp) star $\alpha$~Circini 
was observed over 33 of its 4.479-d rotational periods. The blue and red 
filter light curves were found to be qualitatively different, with the 
blue light curve showing a single-wave variation, whereas the red light 
curve exhibited a double-wave structure. Both were analysed using 
Bayesian photometric imaging. In addition, the strongest roAp-type 
oscillation were recovered in both filters,
demonstrating the photometric sensitivity of 
BRITE data. We refer to \cite{weiss16} for details and the full analysis.


Together with the brightest roAp star, the brightest $\beta$~Cephei 
star, $\beta$ Centauri, was studied during the same observing campaign. 
This triple system contains two pulsating B-type stars, one of them with 
a measured magnetic field. Both stars exhibit pressure and gravity 
modes. Previous studies of the system revealed two oscillation modes 
spectroscopically; photometric studies remained inconclusive. BRITE 
observations revealed a completely different picture. A total of 17 
independent frequencies (8 likely g mode pulsations and 9 p modes) and 
two combinations thereof were detected \cite{pigu16}. The orbital 
light-time effect may reveal which frequencies in the light variations 
originate from which star, but since the orbital period of the B-star 
binary is close to one year, more observations are needed to this end - 
and are certainly worth the effort.

The third of the initial three BRITE science publications reported a 
study of two Be stars, $\mu$ and $\eta$ Centauri \cite{baade16}. Whereas 
in $\mu$ Cen, strong light echoes from the inner circumstellar disk, 
which is viewed face-on, prevented new insights into the pulsational 
properties, the study of $\eta$ Cen turned out to be more revealing. The 
modulation frequency of the star-disk mass transfer is equal to the 
frequency difference of two nonradial pulsation modes. This low 
frequency also modulates the (large) amplitude and the frequency of a 
circumstellar (so-called \v{S}tefl frequency), which is slightly lower 
than those of the stellar pulsations. The modulations in the mass loss 
are linked to variations in this \v{S}tefl frequency. This implies the 
presence of two engines causing the pulsation-mass loss interactions, a 
pulsational one from the star, and one related to viscosity in the 
circumstellar disk. In-depth arguments and discussions can be found in 
\cite{baade16}.

Another $\beta$ Cephei studied by BRITE is $\nu$ Eridani \cite{hand16}, 
observed in the Orion II field by BAb, BLb, BTr and BHr. This star has 
been the subject of detailed observational and asteroseismic studies 
before, which have mostly concentrated on its p-mode spectrum. Earlier 
ground-based observations did detect two low-frequency g modes for $\nu$ 
Eridani, but it was the new BRITE observations that revealed six new 
gravity modes, firmly establishing it as a hybrid pulsator. The 
excitation of these g modes is still a problem for theoretical 
modelling. Whereas it is possible to excite a few g modes in a certain 
frequency range in stellar models, reproducing the whole extent of the 
observed gravity mode domain is extremely difficult. To this end, 
opacity-modified models of $\nu$ Eridani were computed \cite{dasz16}; 
only those can account for the pulsational mode instability. On the 
other hand, such models also have significantly modified interior 
structure and it remains to be seen how realistic these modifications 
are. Future analysis of BRITE measurements of additional $\beta$ Cephei 
stars may provide an answer to this question.

When embarking on new types of observation, and achieving better 
precision than in the past, new scientific discoveries can result. Also 
BRITE has new discoveries to offer. In addition to the discovery of the 
role of difference frequencies in the mass loss of Be stars, BRITE has 
revealed the presence of massive heartbeat stars. These objects are 
close binaries in eccentric orbits, some of which also show tidally 
excited pulsations. The {\it Kepler} mission has discovered 
approximately two dozen such systems (e.g. \cite{thomp12}). However, 
all of these are composed of stars with masses below 2\,M$_{\odot}$. 
With BRITE-Constellation, three more such systems have been discovered 
so far. $\iota$~Orionis \cite{pablo17} consists of an O9-type primary 
($23$\,M$_{\odot}$) and a B-type secondary ($13$\,M$_{\odot}$), where 
the O star also shows tidally induced oscillations and is therefore the 
first of its kind. $\epsilon$ Lupi is the first doubly-magnetic massive 
star binary \cite{shultz15} with components of 8.7 and 7.3\,M$_{\odot}$, 
respectively, and it shows peak flux modulation at periastron 
\cite{wade17}. Finally, $\zeta$ Centauri ($M \approx 7$\,M$_{\odot}$) 
also exhibits a light curve strongly indicative of periastron 
brightening \cite{hand17}; new radial velocities have been obtained to 
confirm this hypothesis.

\section{Conclusions}
\label{sec-con}

Being the first nanosatellite mission carrying out astrophysical 
research, BRITE-Constellation had to overcome many ``childhood 
diseases'', including initially unexpected radiation damage of its CCD 
detectors, interference issues impeding the communication between the 
satellites and the European-based ground stations, and systematic 
effects occurring in the light curves. However, the child has grown up 
and developed antibodies successfully fighting these diseases. The 
mission has been in full science operation for about two years and is 
now in its presumably most fruitful phase of data acquisition.

The BRITE project is open for collaboration that can be established in 
several ways, be it through the submission of target proposals (a call 
is open at the time of writing), be it through joining the BRITE 
International Advisory Team (BIAST) or the Ground-Based Observing Team 
(GBOT), or be it through attending the next BRITE science conference 
which will take place in Canada in August 2017.

For more information on the project itself, visit the BRITE home page 
{\tt http://www.brite-constellation.at/}. Should you be interested in 
BRITE data, have a look at the BRITE photometry wiki ({\tt 
http://brite.craq-astro.ca/doku.php}). In case you are a social media 
user, feel free to connect through the BRITE-Constellation facebook 
page.

\section*{Acknowledgments}

The Polish contribution to the BRITE project is supported by the NCN 
grants 2011/01/B/ST9/05448, 2011/01/M/ST9/05914, 2011/03/B/ST9/02667 and 
2015/18/A/ST9/00578. GAW and SMR acknowledge Discovery Grant support 
from the Natural Science and Engineering Research Council (NSERC) of 
Canada. AFJM is grateful for financial aid from NSERC and FQRNT 
(Quebec). KZ acknowledges support by the Austrian Fonds zur F\"orderung 
der wissenschaftlichen Forschung (FWF, project V431-NBL). BTr operations 
are supported through a Canadian Space Agency (CSA) Academic Development 
grant. RK and WW acknowledge financial support by the Austrian Space 
Application Programme (ASAP) of the Austrian Research Promotion Agency.

\vspace{6mm}

%
%

\end{document}